\documentclass[onecolumn,12pt,aps,prd,nobibnotes,showpacs,showkeys,
superscriptaddress]{revtex4-2}
\usepackage{graphicx,graphics,color,epsfig}
\usepackage[colorlinks=true,linkcolor=blue,citecolor=blue,urlcolor=blue,
breaklinks=true]{hyperref}
\usepackage{amsmath,amsfonts,amssymb}
\usepackage{verbatim}
\usepackage{latexsym}
\usepackage{dcolumn}
\usepackage{bm}
\usepackage{natbib}
\usepackage[english,russian]{babel}
\usepackage[utf8]{inputenc}

\begin{document}
\title{On phenomenological relations for masses and mixing parameters of
quarks and leptons}
\author{V.~V. Khruschov}
\email{khruschov{\_}vv@nrcki.ru}
\affiliation{National Research Center ``Kurchatov Institute'',
Kurchatov~place~1, 123182~Moscow, Russia}
\author{S.~V. Fomichev}
\email{fomichev{\_}sv@nrcki.ru}
\affiliation{National Research Center ``Kurchatov Institute'',
Kurchatov~place~1, 123182~Moscow, Russia}

\begin{abstract}
A comparative analysis of a number of phenomenological relationships between
constants of the extended Standard Model of electromagnetic, strong and weak
interactions of fundamental particles (hereinafter referred to as the extended
Standard Model) is carried out to detect possible correlations between
constants in the quark and lepton sectors. The presence of such correlations
may indicate connections between constants within a more general theory than
the extended Standard Model. Phenomenological relationships between masses of
current and constituent quarks and mixing angles are considered. A
phenomenological relationship between neutrino masses and neutrino mixing
angle is found. The quark-lepton complementarity relationship for quark and
neutrino mixing angles is confirmed. A typical estimate of the accuracy of
such relationships is given. Estimates of the masses of sterile (right-handed)
neutrinos are obtained at the phenomenological level using the seesaw
mechanism. In this case, it is assumed that the values of the active neutrino
masses depend on three characteristic scales, and the sum of the active
neutrino masses is limited from above by $0.06$~eV. An example of a grand
unification theory and possible stages of spontaneous breaking of its gauge
symmetry to the level of gauge symmetry of the extended Standard Model due to
additional Higgs particles are considered.
\end{abstract}



\maketitle

\section{INTRODUCTION}
\label{sec1}

The quantum theory of electromagnetic, strong and weak interactions of
fundamental particles such as quarks (which form hadrons), leptons, gauge
bosons and Higgs bosons is currently the Standard Model (SM) (see, for
example, \cite{1,2}). It contains quantum electrodynamics, quantum
chromodynamics and the quantum theory of weak interactions
(Glashow--Weinberg--Salam theory). The SM predictions for various processes
depend on the values of its constants, which are not specified within the SM
framework but are found phenomenologically from the results of various
experiments. For example, in the extended SM ($\nu$SM) with massive neutrinos
(in the SM neutrinos are massless), the number of constants is generally 28,
while in the original SM there were 19. The $\nu$SM constants include 12 quark
and lepton masses, 10 left-handed quark and neutrino mixing angles and phases,
3 coupling constants of the electromagnetic, strong and weak interactions, the
gauge weak boson mass, the Higgs boson mass and the value of the angle
$\theta_{QCD}$ associated with CP violation in quantum chromodynamics. In
order to use the couplings between some constants, it is desirable that their
values do not change with time or spatial position during the experiments.
Currently, the results of the search for variations of the fundamental
physical constants (FPC) on large space-time scales \cite{4} confirm the
constancy of their values.

However, the $\nu$SM with the above large number of phenomenological constants
cannot be the final fundamental theory, and in the future it will have to be
unified with the theory of gravity, containing at least two additional
constants: the gravitational constant and the value of the constant associated
with the accelerated expansion of the Universe (see, for example, \cite{5}).
One of the possible ways to solve this problem is to move to a grand unified
theory (GUT), which will unite electromagnetic, strong and weak interactions.
After that, a transition to the theory of unified interaction (TEI), which
unites the GUT and the theory of gravity, is possible. As an acceptable
version of the GUT, a theory with the gauge group $SO(10)$ can be chosen. It
was proposed as one of the first GUTs and still remains an acceptable
candidate. Its advantages include the required set of fermions of the
fundamental representation, as well as suitable schemes of spontaneous
symmetry breaking to the level of the SM gauge symmetry, one of which is given
below:
\begin{equation}
SO(10)\to SU(4)_{ec}\times SU(2)_R\times SU(2)_L\to SU(3)_{c}\times
U(1)_{B-L}\times SU(2)_R\times SU(2)_L\to \nonumber %\\
\end{equation}
\begin{equation}
\to SU(3)_{c}\times U(1)_{Y} \times SU(2)_L,
\label{n1}
\end{equation}
where $SU(4)_{ec}$ is the extended color group, which includes the quark
colors and the lepton color \cite{15} (see also \cite{14}); $SU(2)_R$ is the
right symmetry group of superweak interactions; $U(1)_{B-L}$ is the symmetry
group of the difference between the baryon and lepton charges;
$U(1)_{Y}\times SU(2)_L$ is the SM gauge group, which contains the hypercharge
subgroup $U(1)_Y$ and the left symmetries subgroup $SU(2)_L$ of weak
interactions. Each stage of spontaneous symmetry breaking (indicated by the
arrow above) is associated with non-zero vacuum expectation values of the
corresponding Higgs fields, of which, as follows from relation (\ref{n1}),
there can be quite a lot.

The fundamental spinor representation of the $SO(10)$ group includes one
generation of quarks, leptons, and the CP partner of the right-handed
neutrino. Neutrino masses arise naturally in such a theory, and their
existence is confirmed by experimental data \cite{7}. The $SO(10)$ theory
leads to an extended Higgs sector, allowing for the existence of a large
number of Higgs bosons. Note that the intermediate model arising from the
reduction of $SO(10)$ theory to the SM, which is based on the symmetry
$U(1)_{B-L}\times SU(2)_R\times SU(2)_L$, is currently very popular when
considering possible generalizations of the SM in the electroweak sector. This
is the so-called left-right symmetric model for electroweak and superweak
interactions carried by the photon, left and right W and Z bosons. The
fundamental issue in the experimental verification of this model is the
detection of new right W and Z bosons. For example, the estimates made in
\cite{17} lead to a lower limit on the masses of these bosons of the order of
2~TeV, which allows us to study this energy region in the foreseeable future,
in contrast to the region of the order of $10^{16}$~GeV, which, according to
numerous estimates, is the region of realization of unbroken GUT symmetry with
a simple gauge symmetry group. The results of the predictions of the theory
with a simple $SO(10)$ group and a possible comparison with future
experimental data are considered in detail, for example, in \cite{18,19}.

In the sector of gravitational physics, extensive evidence for the existence
of dark matter (DM) is now available. However, the fundamental nature of DM
remains unknown \cite{7}. To explain the nature of DM (see, e.g.,
\cite{ref24}), many candidates for the role of DM components have been
proposed (including some belonging to new physics), such as WIMPs -- weakly
interacting massive particles \cite{ref32,ref7}, ALPs -- axion-like particles
\cite{ref27,ref18}, and PBHs -- primordial black holes
\cite{ref36,ref9,ref25}. A PBH is a hypothetical object that formed in the
early Universe, where the energy density is so high that matter could collapse
into a black hole. Due to the special formation process, PBHs have an
extremely wide range of masses, from the Planck mass of $\sim 10^{16}$~TeV or
$\sim 2\times10^{-5}$~g to $\sim 10^5$ $M_{\odot}$ depending on the formation
time \cite{ref20,ref19}. The lower mass limit for DM particles may be around
$10^{-22}$~eV \cite{7}, while the mass of heavy dark matter (HDM) and
superheavy dark matter (SHDM) particles varies from 1~TeV to the Planck mass
\cite{ref17,ref30,ref35}.

The best candidates for the role of dark-matter particles are WIMPs. The
masses of WIMPs ($M_{wimp}$) are typically in the range from 10~GeV to 10~TeV
(see \cite{ref33}). In this paper we will assume that $M_{wimp}> 10$~TeV
(taking into account the limit for new physics and new particles from
\cite{8}). It is very likely that the masses of fermion WIMPs, such as heavy
right-handed Majorana neutrinos, can be related to very small masses of active
neutrinos according to the seesaw mechanism (see \cite{ref1}). The seesaw
mechanism of generation of masses of left-handed active neutrinos due to
spontaneous symmetry breaking and large masses of right-handed (sterile)
neutrinos is discussed in detail in \cite{ref10} (see also
\cite{ref23,ref13}).

The aim of this paper is to use known and search for new phenomenological
relations between the values of the masses and the mixing parameters of quark
and leptons. It is obvious that if the existence of a single GUT is possible,
then its manifestation in the form of relations between the constants of the
$\nu$SM in the low-energy region is quite realistic. First of all, in Section
\ref{sec2} we will consider the already found phenomenological relations
between the quark masses and the Cabbibo angle of quark mixing, which can
reflect the relations between the constants under study within the GUT. We
will also test the complementarity hypothesis for the quark and neutrino
mixing angles. We will obtain a new phenomenological relation between the
masses and the mixing angle of neutrinos in Section \ref{sec3}. Then, in
Section \ref{sec4} we will estimate the masses of heavy sterile (right-handed)
neutrinos using the seesaw mechanism. In Section \ref{sec5} we will present an
original parameterization of the mixing matrix of three active and three
sterile neutrinos. In conclusion, we will highlight the main results of the
work and note the possible connection between the decay of heavy sterile
neutrinos and the registration of active neutrinos with ultra-high energy.

\section{PHENOMENOLOGICAL RELATIONS FOR MASSES AND MIXING ANGLES OF QUARKS}
\label{sec2}

The mixing angles and phases of left quarks and neutrinos are included in two
mixing matrices: the Cabibbo--Kobayashi--Maskawa matrix $V_{CKM}$ and the
Pontecorvo--Maki--Naka\-gawa--Sakata matrix $U_{PMNS}$, respectively. These
matrices describe the transition from states of particles with a certain mass
to states with a certain flavor that directly participate in interactions.
Both matrices contain a unitary $3\times 3$ matrix $V$, which has the same
form (with its own mixing angles) for quarks and neutrinos in the standard
parameterization \cite{7}:
\begin{equation}
V = \left(\begin{array}{ccc}
c_{12}c_{13} & s_{12}c_{13} & s_{13}e^{-i\delta}\\
-s_{12}c_{23}-c_{12}s_{23}s_{13}e^{i\delta} &
c_{12}c_{23}-s_{12}s_{23}s_{13}e^{i\delta} & s_{23}c_{13}\\
s_{12}s_{23}-c_{12}c_{23}s_{13}e^{i\delta} &
-c_{12}s_{23}-s_{12}c_{23}s_{13}e^{i\delta} & c_{23}c_{13}
\end{array}\right),
\label{V-matrix}
\end{equation}
where $c_{ij}\equiv\cos\theta_{ij}$, $s_{ij}\equiv\sin\theta_{ij}$,
$V_{CKM}\equiv V$, $U_{PMNS}=VP$,
$P=\{\exp(i\alpha_{CP}),\,\exp(i\beta_{CP}),\,1\}$, $\delta\equiv\delta_{CP}$
is the Dirac phase associated with CP violation in the lepton sector, while
$\alpha_{CP}$ and $\beta_{CP}$ are the Majorana phases associated with CP
violation in the neutrino sector.

Some phenomenological relations between quark masses and mixing angles are
known. For example, in \cite{13} a relation was obtained for the Cabibbo angle
(in the scheme with the matrix $V_{CKM}$ this is the angle $\theta_{12}$),
containing the values of the current quark masses, namely:
\begin{equation}
\theta_{12}=\arctan\sqrt{m_d/m_s}-\arctan\sqrt{m_u/m_c},
\label{quarks_mass_mixing}
\end{equation}
where $m_a$ with index $a$ are the masses of the current $d$, $s$, $u$, $c$
quarks. The relation (\ref{quarks_mass_mixing}) can be generalized by
replacing the masses of the current $d$, $s$, $u$, $c$ quarks with the masses
of the constituent $D$, $S$, $U$, $C$ quarks \cite{13a,14,14a}, i.e.
\begin{equation}
\theta_{12}=\arctan\sqrt{m_D/m_S}-\arctan\sqrt{m_U/m_C}.
\label{quarks_mass_mixing_1}
\end{equation}
We will take the values of the constituent quark masses from the results
obtained within the framework of relativistic potential models \cite{14}:
$m_U=(335\pm2)$~MeV, $m_D=(339\pm2)$~MeV, $m_S=(510\pm15)$~MeV,
$m_C=(1580\pm30)$~MeV. We substitute these values into the formula
(\ref{quarks_mass_mixing_1}) and obtain that $\theta_{12}=14.5^{\circ}$, while
the experimental value of this angle is $\theta_{12}=13.00(4)^{\circ}$
\cite{7}. Thus, the obtained values agree at a level of about 10\%. Note that
if we use the hypothesis of quark-lepton complementarity, namely:
$\theta_{12}+\theta_{12}^{'}=45^{\circ}$ (see, for example, \cite{14}), then
the angle $\theta_{12}^{'}$ for the neutrino mixing matrix $U_{PMNS}$ turns
out to be equal to $\theta_{12}^{'}=30.5^{\circ}$ (the experimental value
$\theta_{12}^{'}=33.4(3)^{\circ}$ \cite{7}).

\section{PHENOMENOLOGICAL RELATIONS FOR MASSES AND MIXING ANGLES OF NEUTRINOS}
\label{sec3}

The SM has 9 fewer constants than the $\nu$SM, which contains additional
constants of the neutrino sector, namely, the neutrino masses, mixing angles,
and phase shifts. These additional constants, which are zero in the SM, can be
determined phenomenologically from the characteristics of processes involving
neutrinos. The most accurate data are obtained in neutrino oscillation
processes, in which the differences in the squares of the neutrino masses,
three angles, and the Dirac mixing phase shift \cite{7} are determined. When
using the formula (\ref{V-matrix}) for the $U_{PMNS}$ matrix, $\theta_{ij}$ is
usually replaced by $\theta_{ij}^{'}$ and the phase $\delta$ by $\delta_{CP}$.
The Dirac phase shifts $\delta$ and $\delta_{CP}$ are related to the CP
violation present in the $\nu$SM. Note that the mixing angles and phases of
the matrices $V_{CKM}$ and $U_{PMNS}$ are fundamental constants.

It is known that the problems of both experimental measurement of neutrino
masses and theoretical explanation of the causes of the emergence of neutrino
masses still remain among the main problems of neutrino physics. Within the SM
framework, neutrinos are considered as massless particles and this fact in the
past corresponded to experimental data. However, contradictions between data
on atmospheric and solar neutrinos, on the one hand, and theoretical
calculations, on the other hand, have arisen for quite a long time. These
contradictions were removed after the adoption of the approach of oscillations
of massive neutrinos, including Mikheev--Smirnov--Wolfenstein resonances in
the processes of neutrino interaction with solar matter.

Neutrino oscillations are now supported by convincing evidence from
experiments with atmospheric, solar, reactor, and accelerator neutrinos (see
\cite{ref26,ref14,7,ref39}). In experiments on oscillations, the differences
in the squares of the masses $\Delta m_{ij}^2=m_i^2-m_j^2$ and the neutrino
mixing parameters are measured. However, the absolute values of the neutrino
mass differences cannot be determined in these experiments, as well as the
Majorana or Dirac nature of the neutrinos.

Since only the absolute value of $\Delta m_{31}^2$ has been experimentally
found so far, the neutrino mass values can be arranged in two ways:
$m_1<m_2<m_3$ and $m_3<m_1<m_2$. The first case of neutrino mass arrangement
is called the Normal Hierarchy of the neutrino mass spectrum (NH), and the
second is the Inverted Hierarchy (IH). The values of the phases $\alpha_{CP}$
and $\beta_{CP}$ are currently unavailable from experimental data, as is the
absolute neutrino mass scale, for example, the value of mass $m_1$.

Three types of experimental data are sensitive to the absolute neutrino mass
scale, namely beta decay data, neutrinoless double beta decay data, and some
cosmological and astrophysical data. Each of these data types measures a
certain observable neutrino mass. These observable masses are the mean
neutrino mass $m_C$, the kinematic neutrino mass $m_{\beta}$, and the
effective neutrino mass $m_{2\beta}$, which are given below:
\begin{equation}
\label{mc}
3m_C=\sum_{i=1,2,3}m_i,
\end{equation}
\begin{equation}
\label{mb}
m_{\beta}^2=\sum_{i=1,2,3}|U_{ei}|^2m_i^2,
\end{equation}
\begin{equation}
\label{mbb}
m_{2\beta}=\bigg|\sum_{i=1,2,3}U_{ei}^2m_i\bigg|.
\end{equation}
The effective neutrino mass $m_{2\beta}$ is the upper diagonal element of the
mass matrix for Majorana neutrinos $M=Um^dU^T$, where
$m^d={\rm diag}\{m_1,m_2,m_3\}$. The absolute values of the two additional
diagonal matrix elements $m_{\mu\mu}$ and $m_{\tau\tau}$ are most likely equal
to each other. This assumption is consistent with some models for the neutrino
mass matrix (see \cite{ref34,ref16}).

The currently available parameters of neutrino oscillations (see, e.g.,
\cite{ref39}) at the $1\sigma$ level, which determine the flavor oscillations
of three light active neutrinos, are presented below:
\begin{subequations}
\begin{equation}
\sin^2{\theta_{12}^{'}}= 0.307^{+0.012}_{-0.011}\,,\qquad
\sin^2{\theta_{23}^{'}}=\left\{{\rm NH}:\,\, 0.561^{+0.012}_{-0.015}
\atop {\rm IH}:\,\, 0.562^{+0.012}_{-0.015} \right., \tag{8a}
\end{equation}
\begin{equation}
\sin^2{\theta_{13}^{'}}=\left\{{\rm NH}:\,\, 0.02195^{+0.00054}_{-0.00058}
\atop{\rm IH}:\,\, 0.02224^{+0.00056}_{-0.00057}\right.,\qquad
\delta/^{\,\circ}=\left\{{\rm NH}:\,\, 177^{+19}_{-20}\atop
{\rm IH}:\,\, 285^{+25}_{-28} \right., \tag{8b}
\end{equation}
\begin{equation}
\Delta m_{21}^2/10^{-5}\,{\text{eV}}^2=7.49^{+0.19}_{-0.19}\,,\qquad
\Delta m_{31}^2/10^{-3}\,{\text{eV}}^2=\left\{\!\!{\rm NH}:
\,\, 2.534^{+0.025}_{-0.023}\atop {\rm IH}:\,\, -2.510^{+0.024}_{-0.025}
\right.. \tag{8c}
\end{equation}
\label{dat}
\end{subequations}

The following experimental limits for the observed neutrino masses were
obtained at 90\% CL \cite{ref39}: $m_C<0.013-0.1$~eV \cite{ref15,ref29,ref5},
$m_{\beta}<0.45$~eV \cite{ref6}, $m_{2\beta}<0.028-0.122$~eV \cite{ref3,ref2},
where the last limit should be increased to $0.079-0.180$~eV to account for
the uncertainty in the values of the nuclear matrix elements. Note that the
listed constraints on the observed mass values (\ref{mc}), (\ref{mb}) and
(\ref{mbb}) do not contradict the phenomenological values of the neutrino
masses and NH of their spectrum used in \cite{ZyFomKh,9,10}, namely
$m_1\approx 0.0015$~eV, $m_2\approx 0.0088$~eV and $m_3\approx0.0497$~eV.
Moreover, the results of recent cosmological observations
\cite{ref29,ref5,ref4,ref11,ref22,ref31} testify in favor of the NH mass
ordering variant with a total sum of active neutrino masses of about
$0.06$~eV. Taking into account the current accuracy of experimental values of
mass observables for neutrinos, the masses can be expressed in rounded values
in eV as $m_1=0.001$, $m_2=0.009$, $m_3=0.05$.

By analogy with the relation (\ref{quarks_mass_mixing_1}) for quark masses and
their mixing angles, we will try to find the same type of relation for
neutrino parameters. Direct generalization of (\ref{quarks_mass_mixing_1})
does not lead to the desired result. However, if we assume that the relation
between $V_{CKM}$ and $U_{PMNS}$ contains a transformation from the group of
discrete symmetries that permutes different generations, then we should check
the new relation:
\begin{equation}
\theta_{12}^{'}=\arctan\sqrt{m_{\mu}/m_{\tau}}+\arctan\sqrt{m_2/m_3}
\label{neutrino_mass_mixing}
\end{equation}
This relation holds at the same level of accuracy as the relation
(\ref{quarks_mass_mixing_1}).

\section{THE SEESAW MODEL AND THE MASSES OF RIGHT-HANDED (STERILE) NEUTRINOS}
\label{sec4}

A fundamental question in neutrino physics is the nature of neutrino mass
generation. Lacking a satisfactory theory of this phenomenon, the question can
currently be considered at a phenomenological level. Let us first assume that
there are several different contributions to the neutrino mass, and two of
them are the most important. For example, we can assume that the first
contribution is related to the Majorana mass of the light left-handed
neutrino, which arises outside the SM. This contribution can arise at some
characteristic scale due to the presence of an effective Majorana neutrino
mass term in the Lagrangian when modifying the SM Higgs sector:
\begin{equation}
L'_m=-\frac{1}{2}\overline{\nu}_{L}M_{\nu}\nu^c_{L} + h.c.
\end{equation}
The contribution to the neutrino mass associated with $L'_m$ can be taken into
account through the phenomenological parameter $\xi$. The second contribution
can be associated with the seesaw mechanism that occurs when heavy sterile
(right-handed) neutrinos (HSNs) $N_i$ ($i=1,2,3$) are added to this scheme.
This mass contribution can be written as:
\begin{equation}
M''_{\nu}=-M_D^TM_R^{-1}M_D,
\end{equation}
where $M_D$ is the matrix of Dirac neutrino masses and $M_R$ is the mass
matrix of sterile neutrinos. That is, at least one more new scale appears,
related to the masses of heavy sterile neutrinos. Let us take a value for
$M_D$ proportional to the mass matrix of charged leptons
\cite{ref10,ref13,ref37}:
\begin{equation}
M_D=\sigma m_l,
\end{equation}
where $m_l={\rm diag}\{m_e,\,m_{\mu},\,m_{\tau}\}$, and we set $M_R\sim M$.
Thus, the following phenomenological formula can be used to estimate the
neutrino masses:
\begin{equation}
m_{\nu i}=\pm\xi-\frac{m_{li}^2}{M}.
\label{for}
\end{equation}

Using the obtained experimental data (\ref{dat}) on neutrino oscillations, one
can find the absolute values of the neutrino masses $\mu_i$ and the
characteristic scales $\xi$ and $M$ in eV. For example, in the case of NH
\cite{ref37}:
\begin{equation}
NH: \mu_1\approx0.0693, \; \mu_2\approx0.0698, \; \mu_3\approx0.0851, \;
\xi\approx0.0693, \; M\approx2.0454\times10^{19}.
\label{mnh}
\end{equation}

The estimates of the neutrino masses $\mu_i$ obtained in this way turn out to
be too large and do not satisfy the upper limit for the sums of the active
neutrino masses, equal to $0.06$~eV. Therefore, we simplify the mass formulas
(\ref{for}) in order to obtain other estimates. If the structure of the Higgs
sector $\nu$SM consists only of Higgs doublets, then the value of $\xi$ can be
set equal to zero. Using three values of the masses of heavy sterile neutrinos
$M_1$, $M_2$, $M_3$ and some coefficient $\kappa$ (most likely, the values of
which are of the order of unity), instead of the formulas (\ref{for}) for the
masses of active neutrinos, we will use the simple formulas \cite{RHN}
\begin{equation}
\label{forn}
m_{\nu i}=\kappa\frac{m_{li}^2}{2M_i}.
\end{equation}
Substituting the neutrino mass values $m_i$ given above \cite{9,10} into
equation~(\ref{forn}), we obtain the following estimates for $M_1$, $M_2$,
$M_3$ ($1$~EeV = $10^{18}$~eV):
\begin{equation}
M_1\approx 81.6\kappa \; \text{TeV},\,\, M_2\approx0.6343\kappa \;
\text{EeV},\,\, M_3\approx31.8294\kappa \; \text{EeV}.
\label{mne}
\end{equation}
It is likely that HSNs with masses $M_1$, $M_2$ and $M_3$ can decay in
timescales shorter than the lifetime of the Universe, and therefore they can
currently only be produced by processes at ultra-high energies.

It is also easy to obtain \cite{ref37} the following estimates of the neutrino
mass observables $m_C$ and $m_{\beta}$ in eV for the NH case ($m_{2\beta}$
depends on the unknown phases $\alpha_{CP}$ and $\beta_{CP}$):
$m_C \approx 0.02$, $m_{\beta}\approx 0.01$, which do not contradict the
existing constraints on these mass observables.

\section{MIXING OF ACTIVE AND STERILE NEUTRINOS}
\label{sec5}

Due to the huge difference in the masses of heavy sterile and active
neutrinos, one should expect very small values of the mixing angles between
them. This fact will be taken into account below by using small values of
$\epsilon$ in the parameterization for the mixing matrix, which was used
earlier in papers \cite{9,10} for the (3+3)-model, including three known
active neutrinos $\nu_a$ ($a=e,\mu,\tau$) and three heavy sterile neutrinos:
the sterile neutrino $\nu_s$, the hidden neutrino $\nu_h$, and the dark
neutrino $\nu_d$. The model contains six neutrino flavor states and six
neutrino mass states, and the transition between them is accomplished using a
$6\!\times\!6$ mixing matrix $U_{\rm mix}$, or the generalized
Pontecorvo--Maki--Nakagawa--Sakata matrix $U_{\rm GPMNS}\equiv U_{\rm mix}$.
$U_{\rm mix}$ can be expressed as a matrix product $V\!P$, where $P$ is a
diagonal matrix containing the Majorana CP phases $\phi_i$, $i=1,\dots,5$,
i.e. $P={\rm diag}\{e^{i\phi_1},\dots,e^{i\phi_5},1\}$. We denote the Dirac
CP phases as $\delta_i$ and $\kappa_j$, and the mixing angles as $\theta_i$
and $\eta_j$, where $\delta_1\equiv\delta_{\rm CP}$,
$\theta_1\equiv\theta_{12}$, $\theta_2\equiv\theta_{23}$ and
$\theta_3\equiv\theta_{13}$. As noted earlier, we consider the NH order of the
mass spectrum of active neutrinos $\nu_i$, $i=1,2,3$, and also
$\delta_{\rm CP}=1.1\pi$. In what follows, we will use particular forms of the
matrix $U_{\rm mix}$.

For compactness of the formulas, we introduce the symbols $\nu_b$ and
$\nu_{i'}$ for the sterile left flavor fields and the sterile left mass
fields, respectively. The fields $\nu_b$ with index $b$ contain the fields
$\nu_s$, $\nu_h$, and $\nu_d$, and $i'$ denotes the set of indices $4$, $5$,
and $6$. The overall $6\!\times\!6$ mixing matrix $U_{\rm mix}$ can be
represented through $3\!\times\!3$ matrices $R$, $T$, $V$, and $W$ as follows:
\begin{equation}
\left(\begin{array}{c}\nu_a\\ \nu_b \end{array}\right)=
U_{\rm mix}\left(\begin{array}{c}\nu_i\\ \nu_{i'}\end{array}\right)\equiv
\left(\begin{array}{cc}R&T\\ V&W\end{array}\right)
\left(\begin{array}{c}\nu_i\\ \nu_{i'}\end{array}\right).
\label{eq_Umix}
\end{equation}
We represent the matrix $R$ in the form of $R=\varkappa U_{\rm PMNS}$, where
$\varkappa=1-\epsilon$ and $\epsilon$ is a small value, while $T$ in
equation~(\ref{eq_Umix}) should also be a small matrix compared to the known
unitary $3\times 3$ active neutrino mixing matrix $U_{\rm PMNS}\equiv
U$ ($UU^+=I$). Thus, when choosing the appropriate normalization, active
neutrinos are mixed, as in $\nu$SM, using the
Pontecorvo--Maki--Nakagawa--Sakata matrix $U$. At the current stage of
research, we will restrict ourselves to the minimum number of parameters of
the matrix $U_{\rm mix}$, which allows us to clarify the main properties of
the model.

We choose a matrix $T$ in the form of $T=\sqrt{1-\varkappa^2}\,a$, where $a$
is an arbitrary unitary $3\!\times\!3$ matrix ($aa^+=I$), and then the matrix
$U_{\rm mix}$ can be written in the following form:
\begin{equation}
U_{\rm mix}=\left(\begin{array}{cc}R&T\\ V&W\end{array}\right)\equiv
\left(\begin{array}{cc}\varkappa U&\sqrt{1-\varkappa^2}\,a\\
\sqrt{1-\varkappa^2}\,bU&\varkappa c \end{array}\right),
\label{eq_Utilde}
\end{equation}
where $b$ is also an arbitrary unitary $3\!\times\!3$ matrix ($bb^+=I$), and
$c=-ba$. Under these conditions, the matrix $U_{\rm mix}$ will be unitary, too
($U_{\rm mix}U_{\rm mix}^+=I$). To simplify the form of $U_{\rm mix}$ and
reduce the number of its parameters, we set $b=-a^{-1}$. With this choice of
$b$, the matrix $c$ is equal to the identity $3\!\times\!3$ matrix, i.e. in
this case sterile neutrinos are mixed only with the participation of active
neutrinos. In general, a matrix of the $U_{PMNS}$ type should be chosen as the
matrix $a$, which will increase the number of unknown parameters. Therefore,
in particular, the following matrix $a$ can be used:
\begin{equation}
a=\left(\begin{array}{ccc}e^{-i\kappa} &0 &0 \\0&\cos\eta &\sin\eta \\
0 &-\sin\eta & \cos\eta\end{array}\right),
\label{eq_matricesa}
\end{equation}
where $\kappa$ is the mixing phase between active and sterile neutrinos, while
$\eta$ is the mixing angle between them. To make the calculations more
concrete, one can use the following trial values of the new mixing parameters:
$\kappa=-\pi/2$, $ \eta=\pm 15^{\circ}$, and also constrain the values of the
small parameter $\epsilon$ by the condition $\epsilon\lesssim 10^{-5}$.

The parameterization $U_{\rm mix}$ (\ref{eq_Utilde}), first introduced in
\cite{9}, is much simpler than the frequently used in this context another
parameterization (see, e.g., \cite{na1,na2}). For example, in this paper, when
using the relation $b=-a^{-1}$ and choosing the matrix $a$ in the form
(\ref{eq_matricesa}) under the parameterization (\ref{eq_Utilde}), only three
new parameters $\epsilon$, $\kappa$ and $\eta$ arise, which have a simple
interpretation of the modulus of active neutrino oscillations, the CP phase
and the mixing angle, respectively.

\section{CONCLUSION}
\label{sec6}

If it is possible to create a theory of electromagnetic, strong and weak
interactions more general than the $\nu$SM, it will depend on a small number
of constants and will lead to connections between the constants of the
existing model. This means that it is necessary to search for relationships
between the currently known constants, which is of interest not only for the
theory, but also for the use of the found relationships in practical
applications. Despite the relatively low accuracy of the fulfillment of the
relationships between the $\nu$SM constants considered in this paper (about
10\%), the presence of such relationships indicates possible ways for further
development of the $\nu$SM. As shown in this paper, such connections can be
the observed correlations between the quark mixing angles and the masses of
constituent quarks, as well as between the quark and neutrino mixing angles.
The relation between the quark and neutrino mixing angles may reflect a
fundamental relationship between the Cabibbo--Kobayashi--Maskawa and
Pontecorvo--Maki--Nakagawa--Sakata matrices, which will be preserved in a
future GUT-type theory. The estimated values of $m_C$ and $m_{\beta}$ may have
practical implications for the interpretation of future neutrino data.

In this paper, we consider the known seesaw mechanism as a possible cause of
the emergence of small masses of active neutrinos. A simple formula
(\ref{forn}) was used to estimate the masses of active neutrinos, which
depends on three characteristic scales of HSNs mass values and some
coefficient $\kappa$, the values of which are probably of the order of unity.
The obtained estimates of the HSNs masses allow us to consider them as
particles of heavy fermionic dark matter, which can be produced in processes
at ultra-high energies. Dark matter particles may also include heavy scalar
and vector particles that participated in the formation of the early Universe.
Some of them may continue to participate in processes at the present time. For
example, the appearance of active neutrinos with very high energies, as in the
observed event \cite{na3}, may be associated with decays of the hidden
neutrinos considered in this paper into sterile neutrinos, accompanied by
high-energy active neutrinos.

\end{document}